\documentclass[pra,floatfix,preprint,nofootinbib,eqsecnum,tightenlines,12pt]{revtex4}
\usepackage{epsfig}
\usepackage{bm}
\usepackage{amsmath}
\input epsf
\numberwithin{equation}{section}
\renewcommand{\theequation}{\arabic{section}.\arabic{equation}}

\def\Uc{{\sf U}}
\def\Dc{{\cal D}}

\def\cf{}
\def\df{}

\def\be{\begin{equation}}
\def\ee{\end{equation}}

\begin{document}

\title{Living in a Superposition\footnote{A pedagogical essay.}}

\author{James B.~Hartle}

\email{hartle@physics.ucsb.edu}

\affiliation{Santa Fe Institute, Santa Fe, NM 87501}
\affiliation{Department of Physics, University of California,
 Santa Barbara, CA 93106-9530}

\date{\today }

\begin{abstract}
This essay considers a model quantum universe consisting of a very large box containing a screen with two slits  and an observer (us) that can pass though the slits. We apply the modern quantum mechanics of closed systems to calculate the probabilities for alternative histories of how we move through the universe and what we see. After passing through the screen with the slits,  the quantum state of the universe is  a superposition of classically distinguishable histories.  We are then living in a superposition. Some frequently asked questions about such situations are answered using this model. The model's relationship to more realistic quantum cosmologies is briefly discussed. 
\end{abstract}


\maketitle

\section{Introduction}
\label{intro}

\begin{figure}[t]
\includegraphics[width=6in]{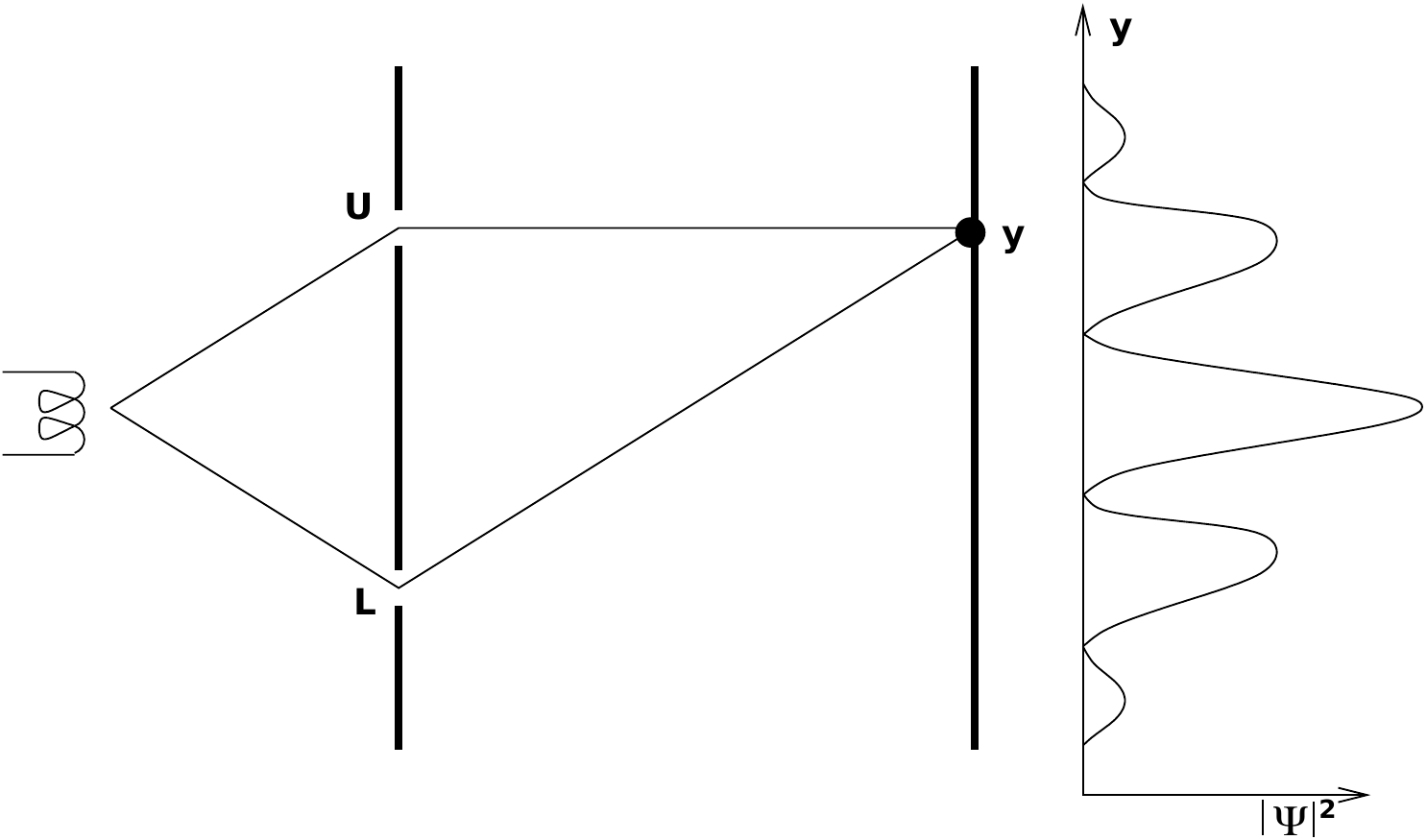}
\caption{The two-slit experiment.  An
electron gun at left emits an electron which is detected at a point $y$ on a screen after passing through another screen with two slits. The interference pattern in arrival positions that appears on the screen is a consequence of quantum interference between a history where the electron went through the upper slit and a history where it went through the lower  slit.  }
\label{twoslit}
\end{figure}

The two-slit thought experiment  sketched in Figure \ref{twoslit}  is the simplest and clearest way of illustrating the phenomena of quantum coherence and decoherence.  The characteristic pattern of places of arrival at the far screen is evidence of quantum interference. Feynman in his basic physics lectures says the two-slit experiment  is ``at the heart of quantum mechanics .... [containing its] only mystery'' \cite{Feylect}.

This pedagogical essay uses a variation of the  two-slit thought experiment to introduce, illustrate and clarify various aspects of the modern formulation of the quantum mechanics of closed systems, most generally the universe as a whole. This formulation is called decoherent (or consistent) histories quantum mechanics (DH) and is the work of many\footnote{For classic expositions at various levels and different emphases see, e.g. \cite{Gri02,Omn94,GH90a,Gel94, Hoh10,QU}. For a tutorial  in the notation used here see \cite{Har93a}.}. In applying quantum mechanics to closed systems like the universe DH can be viewed as an extension, clarification, and to some extend a completion of the work begun by Everett \cite{Eve57}. 

\begin{figure}[t]
\includegraphics[width=4.5in]{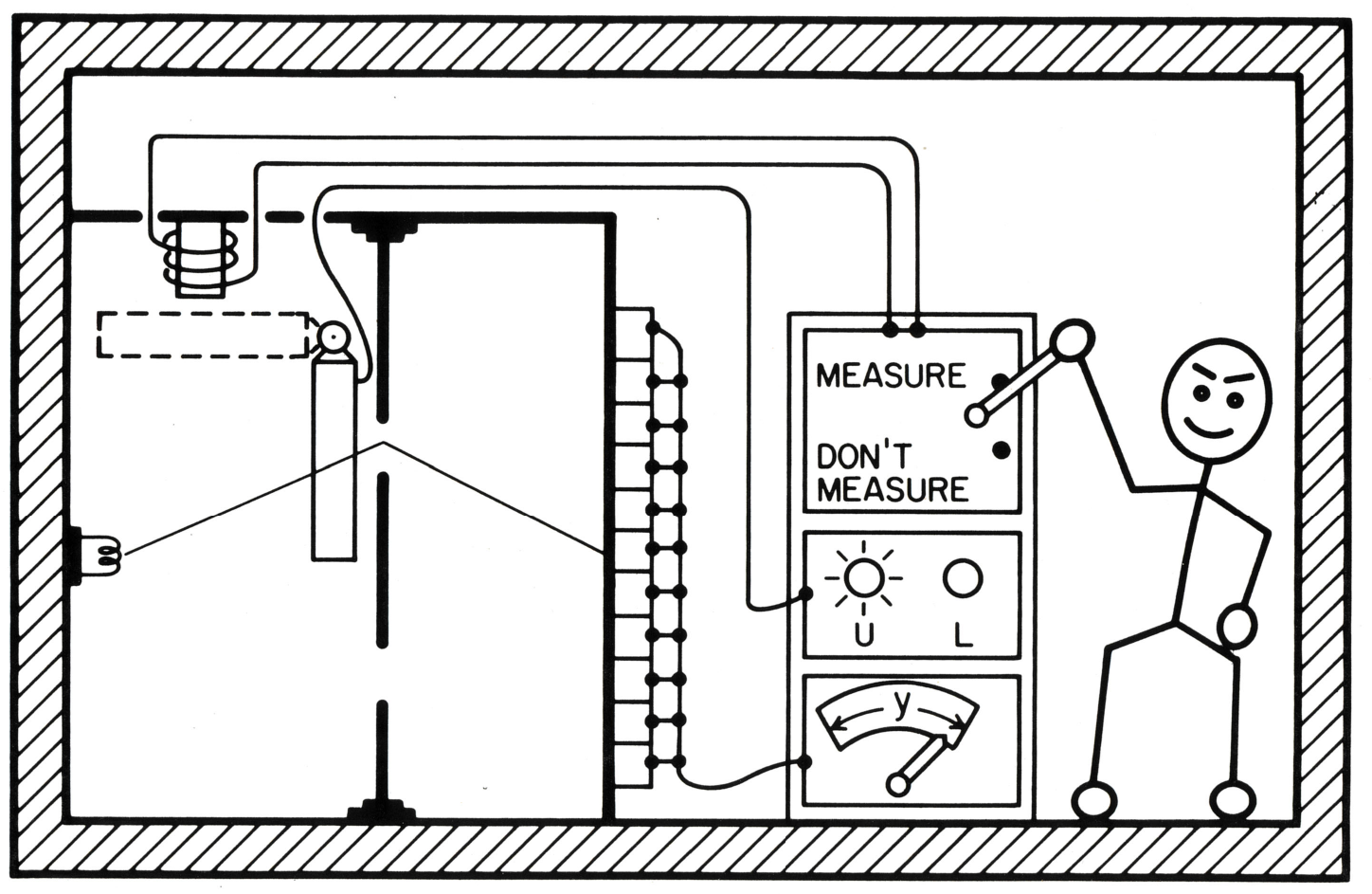}
\caption{A different model closed system. This picture is meant to suggest a closed system that models a laboratory measurement of a two-slit situation like those referred to in Figure \ref{twoslit}. A box contains both a two-slit set-up  and an observer reading the results. The observer is outside the two-slit  subsystem being measured but inside the box.  {\cf Measuring whether the electron went through the upper slit destroys the quantum coherence between  between histories specifying which slit so that probabilities can be assigned to the outcomes of the measurement. }  Such models can also be discussed quantum mechanically but in this paper we are discussing observers that are not outside a two-slit system, but rather inside one.}
\label{twoslitlab}
\end{figure}

{\cf Essential features of DH can already be understood in the context of the two-slit experiment.  The most general objective a quantum theory of a closed system is the prediction of probabilities for alternative histories of how it evolves in time --- probabilities for the history of what happened in the early universe for example. 
But quantum interference is an obstacle to assigning probabilities to sets of alternative histories.  In the two-slit experiment it is not possible to assign probabilities to the alternative histories in which the electron arrives at $y$  having gone through the upper or lower slit. The probability to arrive at $y$ should be the sum of the probabilities of the two histories.  But in quantum mechanics probabilities are squares of amplitudes and  $|\psi_L (y) + \psi_U (y) |^2 \not= |\psi_L (y) |^2 + |\psi_U (y) |^2$ because of interference.  A different physical situation illustrated  in Figure \ref{twoslitlab}  where  the electron interacts with apparatus that
measures which slit it passed through. Quantum interference is destroyed and the set of two histories is said to decohere.  Consistent probabilities can then be assigned to these histories. In a closed system probabilities can be consistently assigned only  to sets of histories that decohere as a consequence of the system's state and Hamiltonian.}

The quantum interference modeled by the two-slit experiment has been seen in many beautiful and important experiments e.g. \cite{Ton89, AHHZ05,hotbucky,JMMetal12,Friedetal00} using increasingly larger interfering systems over time --- molecules more than  a thousand au for example \cite{JMMetal12}.  
Some distinguished scientists expect quantum coherence to break down for sufficiently large systems (e.g. \cite{LG85, Pen96}) but so far there is no evidence for that, and no evidence that quantum interference phenomena could not  be seen for much larger systems. 

Could we, as human observers, be sent through a very large two-slit experiment as in Figure \ref{twoslitpic}? What would we observe as we travel through the screen with the slits and where would we predict we arrive at the farther screen? This paper addresses such questions in the context of DH utilizing a simple model closed system. We are not discussing the feasibility of carrying out such an experiment. We aim rather at a simple thought experiment to illustrate concretely elements of the quantum mechanics of closed systems.

Beyond calculating predictions for our observations we use this example and DH to answer a number of frequently asked questions:  Are we living in a superposition?  If so why don't we see a superposition?  Are we smeared out in space? Is the quantum state reduced when we make an observation? etc. 

As Feynman emphasized, the essential features of the textbook quantum mechanics of measurement outcomes can be illustrated by the two-slit experiment. Here, we shall find that the essential features of the modern quantum mechanics of closed systems (DH) can be illustrated with the  two slit model universe. 

The essay is structured as follows:  Section \ref{model} describes the model two-slit universe with an observer carrying a detector.  The decoherent histories quantum mechanics of this system is described in Section \ref{DH} and used to calculate the predictions for the our  observations in Section \ref{firstandthird}.   The frequently asked questions  mentioned above are discussed and answered in Section \ref{FAQ}.  Section \ref{QC} concludes with a discussion of the relation of this example to realistic quantum cosmology.

\section{A Model Two-Slit Universe}
\label{model}
\begin{figure}[t]
\includegraphics[width=6.5in]{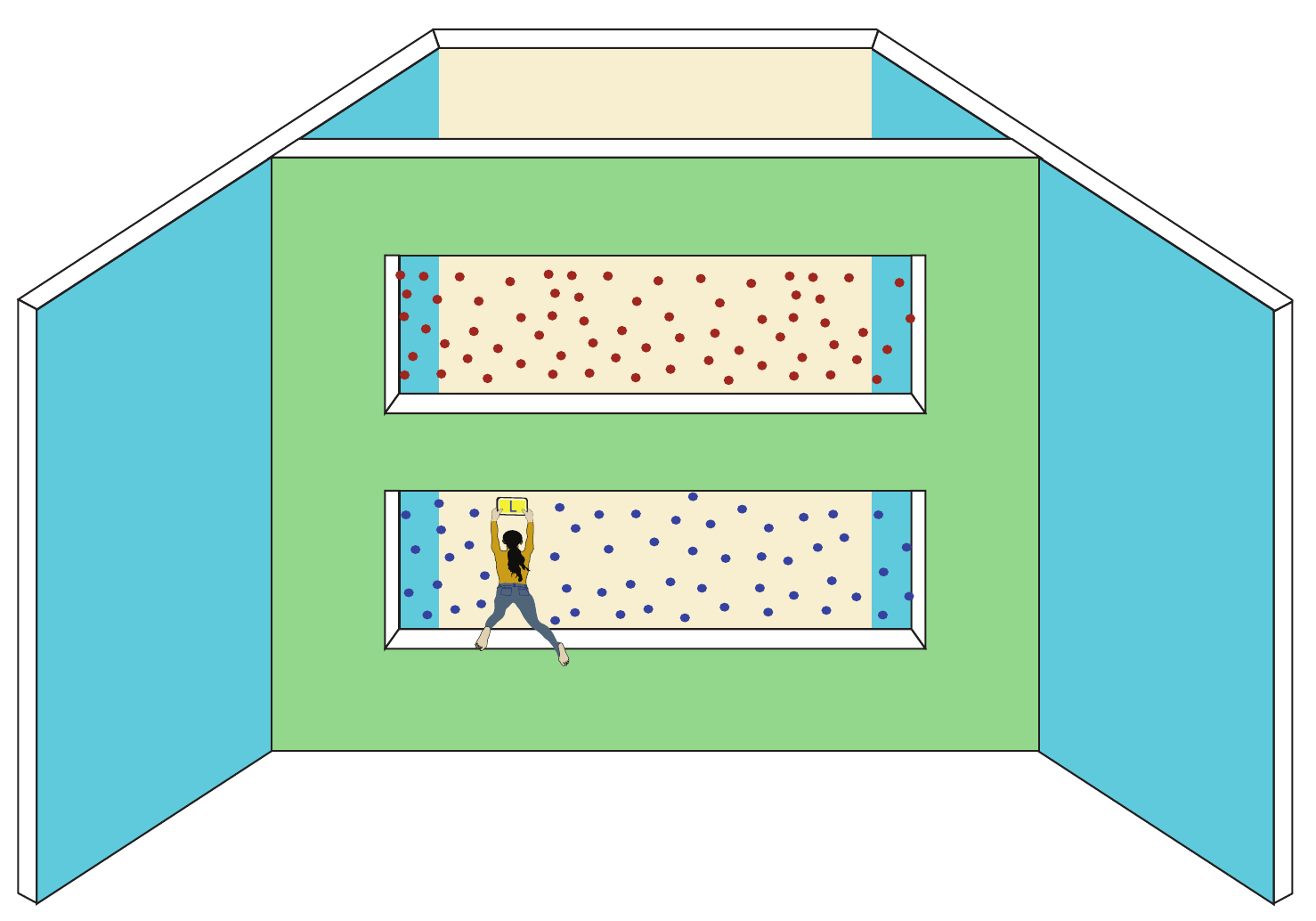}
\caption{A closed system (TSMU) consisting of a very large two-slit apparatus in a closed box together with an observer  (us) carrying a radiation detector that goes through a slit and reaches some point on the farther screen. The top and front of the box have been omitted in the figure to aid visualization  but no photons  or other information are getting in or out through them.  There is a gas of radiation (red) with wavelength $\lambda_U$ near the upper slit, and a gas of radiation (blue) with wavelength $\lambda_L$ near the lower slit. The observer's detector records the presence of radiation and its wavelength.  As pictured, the observer is going through the lower slit recording the blue radiation there with her detector. }
\label{twoslitpic}
\end{figure}

Imagine a large box containing a very large two-slit apparatus as illustrated in Figure \ref{twoslitpic} together with a single observer (us) equipped with apparatus for survival and observation as described more fully below. We stress  that everything relevant for our discussion is inside the box and that nothing outside is interacting with it. We are not considering a  system with an observer outside making measurements on a subsystem as in Figure \ref{twoslitlab}. The box {\it is} the universe. We abbreviate  `two-slit model universe' by TSMU.

We now specify the contents of the box in a little more detail. In the neighborhood of each slit there is a gas of radiation. Near the upper slit the radiation has wavelength $\lambda_U\ll d$ where $d$ is the distance between the slits. 
The radiation near the lower slit has a different wavelength $\lambda_L\ll d$. By measuring the wavelengths on the journey through the slits an observer can tell which slit he or she is passing through\footnote{More fancifully we could imagine the `{\bf U}' and `{\bf L}' are painted near each slit and that the observer is  equipped with a flashlight (torch) and detect which slit by the pattern of reflected radiation.}.

To make these measurements we suppose that the observer is  equipped with detector that can measure which of three states the radiation is in. The three states $\chi_m$  are labeled $m=0,1,2$ having energy levels $0, \hbar/\lambda_U, \hbar/\lambda_L$ respectively. The detector starts its journey in  the ground state $0$. Levels $1$ and $2$ will be excited if it passes through the upper or lower slit respectively. We assume that the observer's center of mass degree of freedom is negligibly affected by either kind of radiation. 

The excitation of the detector creates a record of which slit the observer passed through. If the excitation is to the level $1$ with energy $\hbar/\lambda_U$  then the observer passed through the upper slit, and if to level $2$ with energy $\hbar/\lambda_L$ then it was the lower slit.  If the decay time of these levels is long compared to the travel time to the farther screen then the observer can be said to have a memory of which slit was passed through. That record constitutes a piece of data on which the observer can condition to construct probabilities for further observations as we see in Section \ref{firstandthird}.

Eventually the observer and detector reach the screen at the far end of the box.  To describe their vertical position  there we divide the height up into discrete intervals of length $\Delta$.  These are labeled by a discrete variable $Y$,  $Y= 1,2, \cdots$.  

We next discuss the quantum mechanics of this model.

\section{The Quantum Mechanics of the Model}
\label{DH}

\subsection{The Wave Function of the Universe}
\label{wavu}
\begin{figure}[t]
\includegraphics[width=6.5in]{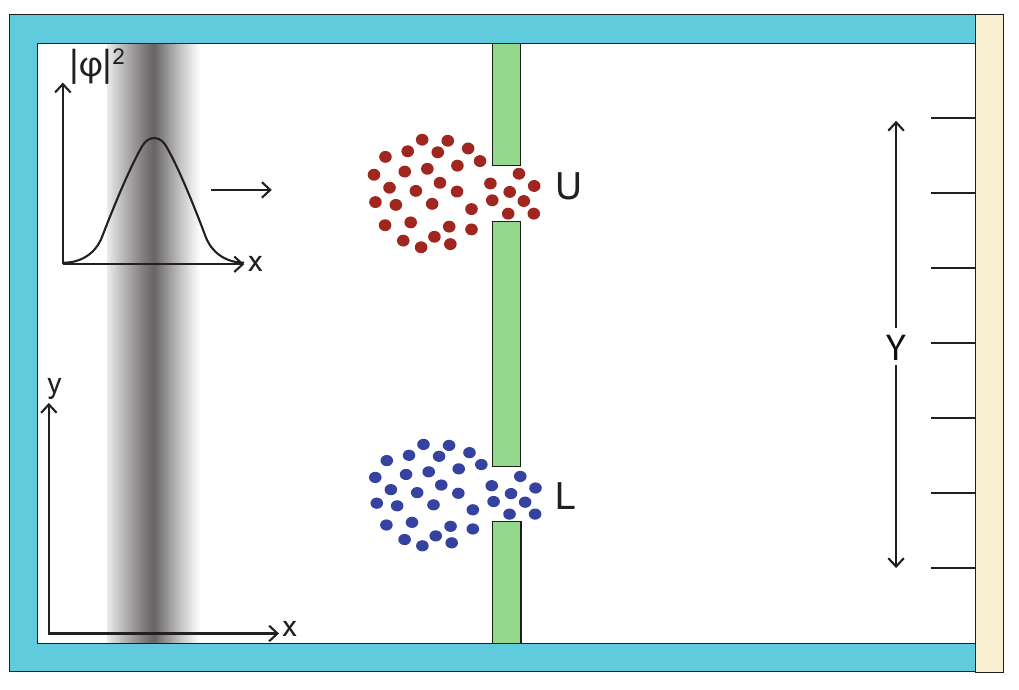}
\caption{A two dimensional slice through the box in Figure \ref{twoslitpic}. This is useful because we assume symmetry in the transverse direction. 
The  slice  is shown at time $t_0$ when the wave function of the observer's center of mass is a narrow wave packet $\phi(x,t)$  in the $x$-direction (inset) (cf \eqref{fullstate}). The packet is moving toward the right so as to arrive  at the slits at time $t_S$ and at the farther screen at time $t_D$. There is a gas of radiation of wavelength $\lambda_U$ near the upper slit (red) and a cloud with a different wavelength $\lambda_L$ near the lower slit (blue).  We consider a set of  alternative histories defined by alternatives at just two times:  Whether the observer went though the upper ($U)$ or lower ($L$) slit at time $t_S$, and what interval $Y$ the observer arrived at the the farther screen at time $t_D$.  }
\label{twoslitu}
\end{figure}

The only degrees of freedom that we follow in this model universe are the center of mass  position of the observer and the state of the radiation detector. The box, slits, screen, radiation, etc are all considered classically. It would be more realistic to consider them quantum mechanically, but we aim at a simple model.  For further simplification we assume symmetry in the direction along the slits. The  model is thus effectively two-dimensional as illustrated in Figure \ref{twoslitu}. The configuration space of the model universe is therefore spanned by the $(x,y)$ coordinates of the observer's center of mass and the three states of the detector $m=0,1,2$. 

The inputs to prediction are the Hamiltonian $H$ and the quantum state of the model universe $|\Psi(t)\rangle$. This is a function of time $t$ in the Schr\"odinger picture in which we work throughout. The state can be described by a wave function on configuration space, viz.
\be
\label{wavu1}
\Psi = \Psi(x,y,m,t).
\ee
We move back and forth between wave functions like \eqref{wavu1} and the representation in terms of bras and kets like $|\Psi(t)\rangle$ as convenient. 

It is important to note that the observer's degrees of freedom and the registrations of the apparatus are inside the closed system not outside it. Their degrees of freedom are arguments of the wave function just like any others.

At the starting time $t_0$ we take the wave function of the  model universe to be
\be
\label{fullstate}
\Psi(x,y,m,t_0) =\phi(x,t_0)\delta_{0m} , \quad t_0<t<t_S.
\ee
where  $\phi(x,t_0)$ is a narrow wave packet peaked to the left of the slits but moving to the right so as to reach the slits at time $t_S$ and the detecting screen at $t_D$ as shown in Figure \ref{twoslitu}. For simplicity we assume that there is no $y$ dependence of the initial wave function over the height of the box.  Thus, progress in $x$ recapitulates evolution in time. The wave function evolves in time by the Schr\"odinger equation
\be
\label{schrod}
i \hbar \frac{\partial \Psi}{\partial t} = H \Psi
\ee
where the Hamiltonian $H$ describes the evolution of the observer's  center of mass position in the presence of the impenetrable walls and the interaction of the detector with the radiation. We won't solve this equation explicitly but rather posit the plausible forms of the solutions.

After passing through the slits the wave function of observer and detector has the form 
\be
\Psi(x,y,m,t)= \psi_U(y,t)\phi(x,t)\delta_{m1}  + \psi_L(y,t)\phi(x,t)\delta_{m2},  \quad t_S<t<t_D .
\label{psiU}
\ee
Here, in the first term $\psi_U(y,t)$ is localized near the upper slit at time $t_S$ and spreads over a larger region of $y$ by the time $t_D$ that the observer hits the detecting screen. The detector is excited to the level $1$ at time $t_S$ and remains in that state of excitation for the rest of the journey to the screen at time $t_D$. Similarly for the second term.

\subsection{Histories}
\label{histories}

\begin{figure}[t]
\includegraphics[width=6.5in]{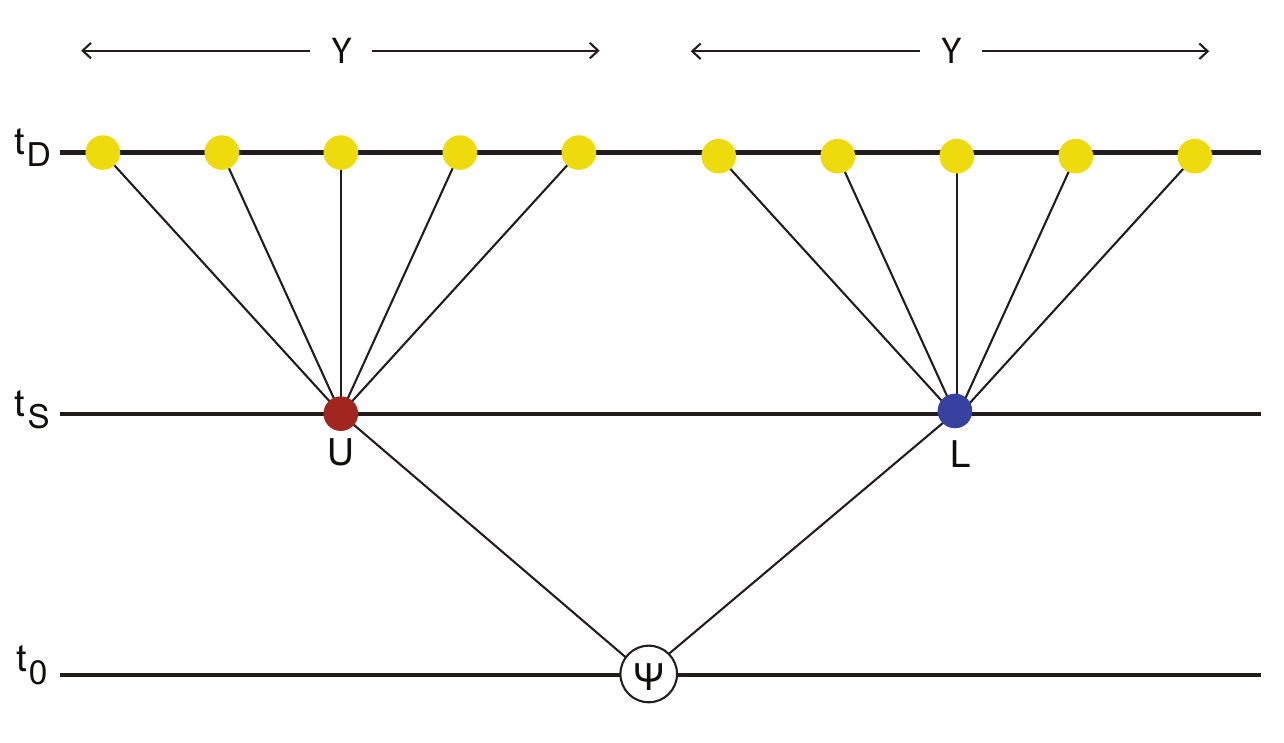}
\caption{A traditional representation of the set of alternative histories discussed in this section. Time runs upward. Each path from $t_0$ to $t_D$ corresponds to a history in which the observer and detector  went through one slit $S$ ($U$ or $L$) at time $t_S$ and arrived at one interval $Y$ at time $t_D$. The branching structure of the set is evident. The initial wave function $\Psi$ divides into further branch wave functions at each step. When all the branch wave functions are mutually orthogonal the set of histories decoheres. }
\label{branchespic}
\end{figure}

Suppose that the observer and detector arrive at position interval $Y$ the screen. What are the probabilities that they went through the upper slit ($U$) or lower slit ($L$)  on the way there?  These are each the probability of a history --- a sequence of events at a series of times. In this case there are just two times ---   the time  $t_S$ that the observer and detector reach the slits, and and the time $t_D$ that it they reach the screen.  The individual histories in an exhaustive set of alternative histories are labeled by $(Y,S)$ where $S$ is $U$ or $L$.

A set of histories like $\{Y,S\}$ has a branching structure illustrated\footnote{For the experts: In this kind of diagram the dots correspond to projection operators representing the various alternatives at different times. The lines connecting the dots and the state represent unitary evolution. The branch state vector of a particular history can be generated from the initial by a sequence of unitary evolutions and projections with a result like \eqref{branches}.} in Figure \ref{branchespic}. A wave function called a `branch wave function' can be associated with each history. The initial wave function divides into two branch wave functions at time $t_S$ and each of those divides into branches for different  $Y$'s at time $t_D$.   The initial wave function is a sum of all the branch wave functions. As we will discuss in the next section, when it is consistent to assign probabilities to a set of histories their probabilities are norms of corresponding branch  wave functions. 

There is a general procedure for constructing branch wave functions from projection operators at representing alternatives at different times. But the set $\{Y,S\}$ is so simple we can read the results off of \eqref{psiU}. The branch wave functions $\Psi_{YS}(x,y,m,t)$ are the two parts of \eqref{psiU} restricted to the various values of $Y$.

The restriction to the intervals $Y$ can be made more explicit by introducing projection operators $P_Y$ that are 1 for $y$'s inside the interval $Y$ and zero for $y$'s outside it.  These satisfy
\be
\label{Yprojections}
P_YP_{Y'}= \delta_{YY'}, \quad \sum_Y P_Y=I 
\ee
showing that they are an exclusive and exhaustive  set of projections. 

The branch wave functions for the set $\{Y,S\}$ at the last time in the histories $t_D$  are then, from \eqref{psiU}, 
\begin{subequations}
\label{branches}
\begin{align}
\Psi_{YU}(x,y,m,t_D) = P_Y \psi_U(y,t_D)\phi(x,t_D)\delta_{m1}   ,    \\
\Psi_{YL}(x,y,m, t_D) = P_Y \psi_U(y,t_D)\phi(x,t_D)\delta_{m2} .
\end{align}
\end{subequations}
These branch wave functions represent branch state vectors which we write $|\Psi_{Y,U}(t_D)\rangle$ and $|\Psi_{YL}(t_D)\rangle$. From \eqref{Yprojections} and \eqref{psiU} we have 
\be
\Psi(x,y,m,t_D) = \sum_{Y,S} \Psi_{YS}(x,y,m,t_D) .
\label{sumbranches}
\ee
The wave function of the model universe is the sum of all its branches. We now turn to the probabilities of these histories.

\subsection{Decoherent Histories and Probabilities}
\label{decoh-probs}

The natural candidates for the probabilities of the history $(Y,S)$ would be square of the norms of its branch wave functions \eqref{branches}, viz
\be
 \label{probs}
 p(Y,S) = || \ |\Psi_{YS}(t_D)\rangle \ ||^2 .
 \ee
By `norm' here we mean the usual inner product between states represented by wave functions. For two states $|\Phi''(t )\rangle$ and $|\Phi'(t)\rangle$ with wave functions $\Phi''(x,y,m,t)$ and $\Phi'x,y,m,t)$ we define
\be
\label{prod}
\langle \Phi''(t)|\Phi'(t)\rangle \equiv  \sum_m \int dx dy \Phi^{''*}(x,y,m,t)\Phi'(x,y,m,t) .
\ee
The norm of a state $|\Phi(t)\rangle$ is then
\be
\label{norm}
||\ |\Phi(t)\rangle \ ||^2 \equiv \langle \Phi(t) | \Phi(t)\rangle .
\ee

The simple example in the Introduction shows that the probabilities  \eqref{probs} are not consistent with the rules of probability theory unless the quantum interference between the branches vanishes. 
More generally the set of histories must decohere to have consistent quantum probabilities. 

A  set of histories decoheres if the branch state vectors  are orthogonal for different histories. For the set $\{Y,S\}$ this is
 \be
 \label{decoherence}
 \langle \Psi_{Y''S''}(t_D) |\Psi_{Y'S'}(t_D)\rangle \propto \delta_{Y''Y'}\delta_{S''S'} .
 \ee
 for all values of $Y$ and $S$.  
This orthogonality is the natural notion of the absence of quantum interference between the branches. 
 
 It is easy to see  from \eqref{branches} giving the branch wave functions that the set $\{Y,S\}$ is decoherent.
 The branches are orthogonal in $Y$ because $P_{Y''}P_{Y'}=0$ for differing $Y''$ and $Y'$ cf \eqref{Yprojections}. The branches are orthogonal in $S$ because the detector states $\chi_m$  are orthogonal. In the next subsection we will check the resulting consistency of the probabilities \eqref{probs} explicitly. 
 
 \subsection{Consistency}
 \label{consist}
 
\begin{figure}[t]
\includegraphics[width=6.5in]{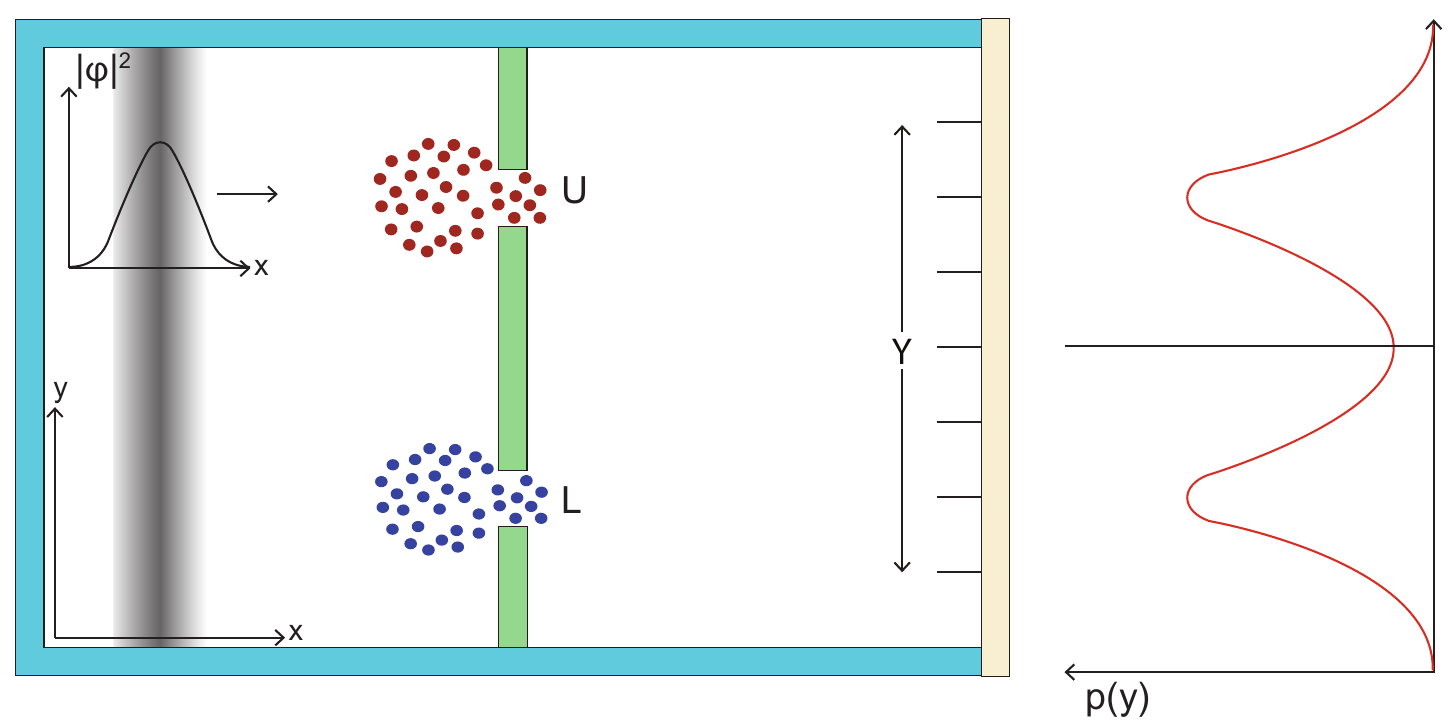}
\caption{A qualitative representation of the probability $p(Y)$ for the observer and detector to arrive in the position interval $Y$ on the end screen.  Because there is a record of which slit the system passed through, the set of histories decoheres, and the probability $p(Y)$ is the sum of the probabilities to go through the upper slit and arrive at $Y$ and the lower slit and arrive at $Y$. The result is not an interference pattern. }
\label{probY}
\end{figure}

 Suppose we ask just for the probability $p(Y)$ that the observer arrives at the interval $Y$ on the far screen at time $t_D$.
The branch state vector for this history is cf \eqref{branches} 
\be
|\Psi_Y(t_D)\rangle= P_Y |\Psi(t_D)\rangle .
\label{branch3}
\ee
Evidently from \eqref{psiU} and \eqref{branches} we have 
\be
\label{branch4}
|\Psi_Y(t_D)\rangle =|\Psi_{YU}(t_D)\rangle +|\Psi_{YL}(t_D)\rangle. 
\ee
The probability to arrive at $Y$ is 
\be
\label{sumrule}
p(Y)=||\ |\Psi_Y(t_D)\rangle||^2  =|| \ |\Psi_{YU}(t_D)\rangle||^2 + | \ |\Psi_{YL}(t_D)\rangle||^2.
\ee
There is no interference term because the two branch state vectors are orthogonal \eqref{decoherence}. 
The probability distribution just for $Y$ is  then just the sum of the probabilities for these two branches these two branches
\be
p(Y)=p(Y,U)+p(Y,L)
\label{pY}
\ee
and shown schematically in Figure \ref{probY}.  Decoherence ensures that probabilities are consistent with usual rules of probability theory.

\subsection{Records}
\label{records}

This set of histories decoheres because a record of which slit the observer passed through is created in the state of the detector.  This association between records and decoherence is a very general one and a fundamental property of DH.  Decoherence implies the existence of records somewhere in the closed system although they may not be in  a form that is accessible or useful to us.  Conversely the existence of records of histories implies their decoherence as in this example\footnote{For the experts we are talking about medium decoherence and strong records.}. Indeed in realistic situations many records of the same history will be created (e.g. \cite{RZZ13}). 

Note that the set of histories  decoheres whether or not the observer looks at the state of excitation. 
The states $\chi_1$ and $\chi_2$ are still orthogonal\footnote{This is the origin of the injunction in Copenhagen quantum mechanics to sum probabilities for alternatives that `could have been measured but were not' \cite{Har93b}.} and \eqref{decoherence} still holds.

The association of decoherence and records has been beautifully demonstrated in the experiment of Hackerm\"uller et al \cite{hotbucky}: A heated buckyball ($C_{60}$) is passed through a Talbot-Lau interferometer exhibiting a quantum interference pattern.  As the temperature is increased the interference pattern dissappears when a temperature is reached where the wavelength of the radiation is short enough that it would contain a record about which slit the buckyball passed through. That is the creation of a record by emission in contrast to the creation by absorption in this toy model. But both bear out the idea that what histories that decohere are histories that are recorded.  

\subsection{Coarse-grained Histories}
\label{cg}
The set of histories $\{Y,S\}$ illustrated in Figure \ref{branchespic} is coarse grained.  The position of the observer is  followed only at two times $t_S$ and $t_D$ and not at all times. And these positions are not followed to arbitrary accuracy but only to the widths of the slits and of the intervals $Y$. The positions in between are unspecified --- a coarse-grained description of the histories.  Specifying only the interval of arrival $Y$ at the farther screen would be a coarser-grained description. Specifying more details of the path taken in between would be a finer-grained description. A coarse graining of a decoherent set like $\{Y,S\}$ is again decoherent.  But fine-graining risks losing decoherence.  Some coarse graining is essential for decoherence.   A completely fine grained description would not decohere.

\section{Third Person and First Person}
\label{firstandthird}

In the quantum mechanics of a closed system like TSMU it is useful to distinguish between two kinds of description of the system and correspondingly two kinds of probabilities \cite{HS09,HH15b}.  

{\it Third Person Descriptions and Probabilities:} Descriptions  of what the universe contains and how that evolves --- histories of what occurs.  Since observers are physical subsystems within the closed system, third person descriptions include a description of the histories of what observers see and how they behave.   All of the previous discussion has been about third person histories of one observer and its detector --- which slit it goes through, how the detector was excited, where it arrives at the screen, etc. The probabilities for these third person histories are called third person probabilities.  They are what is supplied directly by the quantum state of the system and the dynamics. Examples are the probabilities $p(Y,S)$ in \eqref{probs}.

{\it First Person Descriptions and Probabilities:}
Suppose that TSMU is our universe and we are the observer in it. We are interested in the first person probabilities for what {\it we} observe --- what wavelength radiation {\it we} detect,  which $Y$ {\it we}  arrive at etc.  Since we are a physical system within the universe first person probabilities can be derived from the third person probabilities for its histories\footnote{In other papers we have called first and third person probabilities top down and bottom up probabilities respectively e.g. \cite{HH06}.}.  We don't observe whole four-dimensional histories, but rather limited features of the universe from an observing situation that is  localized in space and time. The first person probabilities for what we observe are necessarily conditioned on the data $\Dc$ describing that observational situation including the information about when the observation was made.  The first person probability for an observable $\cal O$ is
\be
\label{1stpers}
p^{(1p)}({\cal O}) =p({\cal O}|\Dc) .
\ee
Here is an example: Our detector makes a transition to level $1$ at time $t_S$. We then know that we have just passed through the upper slit $U$. That is the data $\Dc$. Given that data what do we predict for the probability that we will arrive at position $Y$ on the screen? This is the first person probability
\be
\label{1stpersonY}
p^{(1p)}(Y) =p(Y|U) \equiv p(Y,U)/p(U).
\ee
Using \eqref{probs}  this is
\be
\label{condprob}
p^{(1p)}(Y) = \frac{|| \ |\Psi_{YU}(t_D)\rangle ||^2 }{ || \ |\Psi_U (t_D) \rangle ||^2} .
\ee
This distribution looks like the top part of the graph of $p(Y)$ vs. $Y$ in Figure \ref{probY} but renormalized so that the total probability for some $Y$ is unity.

\begin{figure}[t]
\includegraphics[width=6.5in]{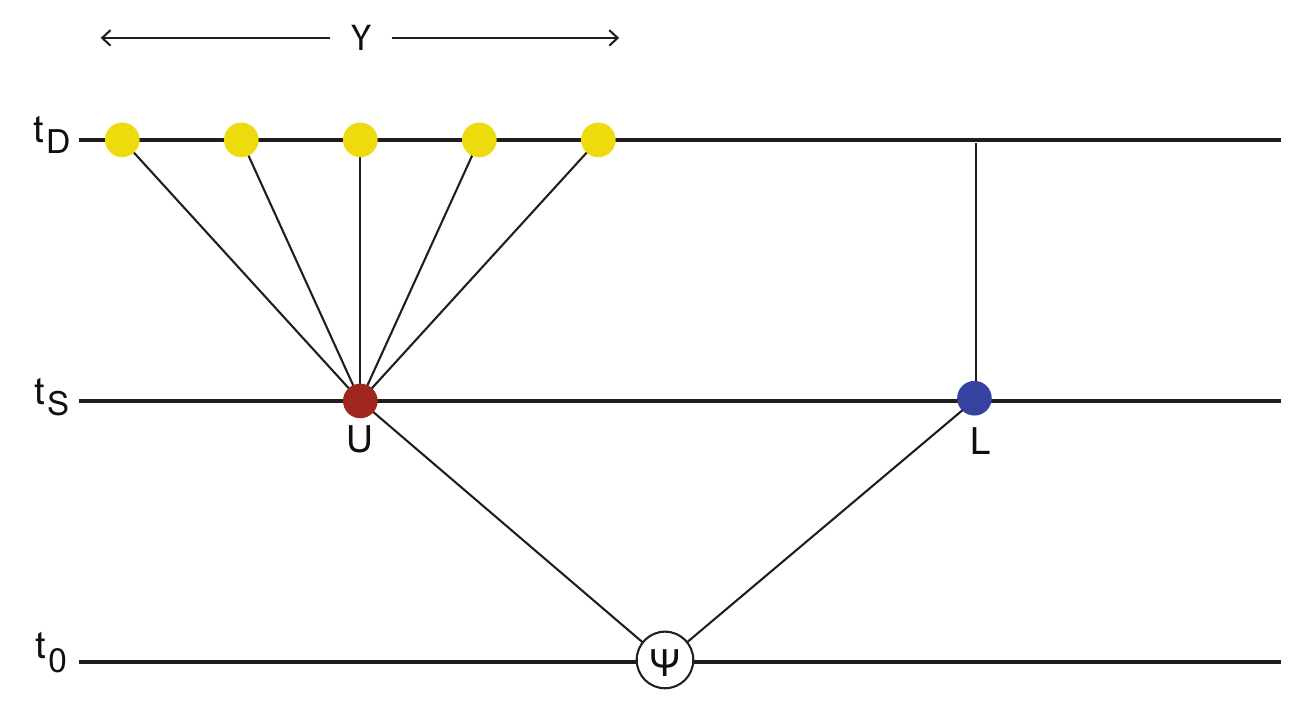}
\caption{There are different third person descriptions of TSMU at different levels of coarse graining. Suppose at time $t_S$ our detector shows we went through the upper slit $U$. We want to predict where we will hit screen a the later time $t_S$. We could calculate using the set in Figure \ref{branchespic}, but it would be inefficient because that also describes  what {\it would} have happened if we went through the lower slit.   A better set of alternative histories of TSMU is the one shown above  that follows which $Y$ is reached only if we go through the upper slit. That is a coarser grained set than the one in Figure \ref{branchespic}.}
\label{branches-adapt}
\end{figure}

{\df To evaluate \eqref{condprob} for $p(Y|U)$ its obviously not necessary to compute the probability of which $Y$ would have been arrived at if we had gone through the lower slit. We only need a set of histories that follows our future observations and ignores (coarse grains) over features of TSMU that are irrelevant for that.  Figure \ref{branches-adapt} shows an example\footnote{ This is an example of an adaptive branch dependent set of histories \cite{GH13}. Different branches at $t_S$ branch  differently in a way that is adapted to what our data are then.}.  

The branch wave functions of the histories in Figure \ref{branches-adapt}  are (cf. \eqref{branches})
\begin{subequations}
\label{branches1}
\begin{align}
\Psi_{YU}(x,y,m,t_D) &= P_Y \psi_U(y,t_D)\phi(x,t_D)\delta_{m1}   ,   \\
\Psi_{L}(x,y,m, t_D) &=  \psi_U(y,t_D)\phi(x,t_D)\delta_{m2} .
\end{align}
\end{subequations}
These are  decoherent because of the orthogonality of the $P_Y$ and the  $\chi$'s.

\section{Some Frequently Asked Questions}
\label{FAQ}
Imagine that we are the observer in TSMU.  A number of questions arise about quantum mechanics in this context. Many of these FAQ's are not clearly defined.  We mostly focus on FAQ's that can be reformulated so that they are answered by appropriate quantum probabilities\footnote{For FAQs that are more about the author's opinion on issues that come up in quantum mechanics like which histories are real, see e.g \cite{Har06}. }. 

\begin{itemize}

\item{\it Are we living in  a superposition?} {\bf Yes.}   Any state can be expressed as a superposition of  other states for example by using a basis.  This FAQ is therefore not precisely defined.  Probably the question of interest  to many is rather whether the state is a superposition of different histories that each can be `classically described'  and are `macroscopically distinct'.  In TSMU  after the time $t_S$ the state of the universe has evolved to the superposition of the two branches in \eqref{psiU}. Each of these can described in classical terms and  each could be said to constructed from alternatives that are `macroscopically distinct'. As the observer in the box,  we are living in such a superposition.  {\cf Schr\"odinger's cat  also lives in a superposition.}

\item{\it  If we are living in a superposition why arn't we  smeared out?}  {\cf The probability for the observer to be in two places at once is zero because the operators vanish that would represent this kind of situation. For example \eqref{sumbranches} shows that  the wave function of the universe is in a superposition of different arrival intervals $Y$. But since $P_Y P_{Y'}=0$ for $Y\ne Y'$ \eqref{Yprojections} it predicts zero probabilities for the observer to be in two intervals at once.  It's the same for the slits.}

\item{\it If we are living in a superposition why don't we see a superposition or feel superposed?} The detector could be part of the observer's brain. Registrations of the detector then  model a physical realization of  `see' and `feel'. We only feel going though the upper slit or  alternatively  feel  going through the lower slit even though the quantum state is a superposition of the two. These are exclusive alternatives. 
The reason we don't see the superposition is that {we are not somehow outside the universe observing  whether its state is a superposition of terms. We are inside the universe participating in one of the terms of the  superposition.} 

\item{\it Does TSMU model a measurement?:}   {\bf Yes.} There is no precise definition of `measurement' in  textbook (Copenhagen) quantum mechanics or in DH.  But the  author would informally characterize the excitation of the detector by the radiation near the slits as `a measurement situation'. A variable --- the wavelength of the radiation --- becomes correlated with an excited state of the detector that can be read by the observer.
Certainly TSMU has many similarities with classic measurement models \cite{LB39} for instance the one in Figure \ref{twoslitlab}. But it also differs from these models in that what is measured is not fixed, but rather determined by the quantum accident of which slit the observer passed through.  Specific measurement situations can be described in DH, and their outcomes predicted, but a precise general notion of measurement is not needed for DH's formulation. 

\item{\it Do the predictions of DH differ from those of textbook quantum mechanics.} 
 {\bf Yes and No.} Textbook quantum theory predicts the probabilities for the outcomes of measurements carried out by one subsystem of the universe on another. DH will predict the same probabilities to an excellent approximation. In TSMU DH would yield the same probabilities for the detector's measurements  of the radiation.  But DH also predicts  probabilities for which measurements are carried out  where,  and for the motion and fate of the observer --- alternatives not considered by textbook quantum theory.  Textbook quantum theory  is not an alternative to DH but rather contained within it as an approximation appropriate for measurement situations. 

\item{\it Would anthropic reasoning modify the predictions of TSMU?} \  {\bf No}. Anthropic reasoning is automatic in DH through the first person probabilities for observation (Section \ref{firstandthird}) \cite{HH15b}.  Suppose that the radiation at the upper slit were intense enough to kill the observer. The third person probabilities for the histories $(Y,S)$ of the motion of the observer and the  registration of the detector would be unchanged. But the first person probabilities for our observations would be affected assuming that the data $\Dc$ contained information that we are alive at the farther screen. The first person probability would be unity that we passed through the lower slit. There is zero first person probability to observe the red radiation which is where we cannot exist. 
 
 \item{\it Is the quantum state  of the universe ever reduced?:}  
 {\bf  No.} In Section \ref{firstandthird} we derived the first person probability $p(Y|\Dc)$ to arrive at an interval $Y$ on the far screen given data $\Dc$ about the detector registration.  Suppose these data imply that we passed through the upper slit $U$. Equation \eqref{condprob} for this probability can be rewritten as
 \begin{subequations}
 \label{reduce}
 \be 
 p(Y|U) = ||P_Y \Uc(t_D,t_S) |\Psi_U(t_S)\rangle||^2 
 \ee
 where
 \be
 \label{reduced}
 |\Psi_U(t_S)\rangle \equiv \frac{P_U |\Psi(t_S)\rangle}{||P_U |\Psi(t_S)\rangle||} .
\ee
\end{subequations}
and $\Uc(t_D,t_S)$ represents unitary evolution from $t_S$ to $t_D$ by the Schr\"odinger equation.
Superficially the formulae \eqref{reduce} are like those in text book quantum mechanics describing the reduction of the state of a subsystem  that occurs when the subsystem undergoes an `ideal' measurement by another subsystem outside it. This was von Neumann's second law of evolution\footnote{This second law of evolution is itself problematical since almost no realistic measurements are `ideal'.}\cite{vNeu32}. 
But this resemblance is misleading. The states and operators in these equations are not of a subsystem of the universe being measured, but rather of the whole thing. The Hilbert space includes both what is observed and the system observing it. In the quantum mechanics of the  universe there is no `other measuring system' and no `second law of evolution'.  Eq \eqref{reduced} is not some mysterious feature of a measurement process. Rather it is  but a step in the construction of conditional probabilities. It is no different from the `reduction' that occurs in horse racing when a particular horse wins and the probabilities for further races which are conditioned on that event become relevant.

\item{\it Does some physical process cause the wave function to branch?}   {\bf No.} {\cf The branching structure in Figure \ref{branchespic} is a {\it choice} of how to describe what goes on in TSMU. Many other descriptions are possible leading to different branching structures. Consider for example the set of histories where only $Y$ is specified and the question of which slit we go through ignored.  Then there would be no branching at $t_S$. Or consider the set of histories in Figure \ref{branches-adapt} where there is no branching after going through the lower slit. DH does not prefer one decoherent set of histories over any other.  All are in principle available to be used by us in the process of prediction although some will be more useful to human observers than others.}

\item{\it Is coarse-graining necessary?}  {\bf Yes.}  {\cf Some coarse graining is necessary for decoherence except in trivial cases. Realistic mechanisms of decoherence involve coarse graining. In the classic example of Joos and Zeh \cite{JZ85} the histories of the positions of a mm size dust grain deep in intergalactic space decohere because of the vast number of CMB photons that scatter from it every second. 
A decoherent set of histories  follows the positions of the dust grain and ignores (coarse grain over) the photons. More generally `environmental decoherence' results from separating a closed system into a subsystem and an `environment' and then coarse graining over its environment (e.g. \cite{GH13}). 
It is a remarkable fact that in the quantum mechanics of closed systems some information must be sacrificed in order to have interesting probabilities at all. } 

\end{itemize}

\section{Quantum Cosmology}
\label{QC}

It is an inescapable inference from the physics of the last century that we live in a quantum mechanical universe. We perhaps have little evidence of peculiarly quantum mechanical phenomena on large and even familiar scales, but there is no evidence that the phenomena that we do see cannot be described in quantum mechanical terms and explained by quantum mechanical laws. If this inference is correct, then there must be a description of the universe as a whole and everything in it in quantum mechanical terms. The nature of this description and its predictions for observations are the subject of quantum cosmology. 

The two-slit universe of this paper is a toy model to illustrate a few aspects of realistic quantum cosmology. We now describe some of the connections between the model and the ongoing program of quantum cosmology.

\begin{figure}[t]
\includegraphics[height=3.5in]{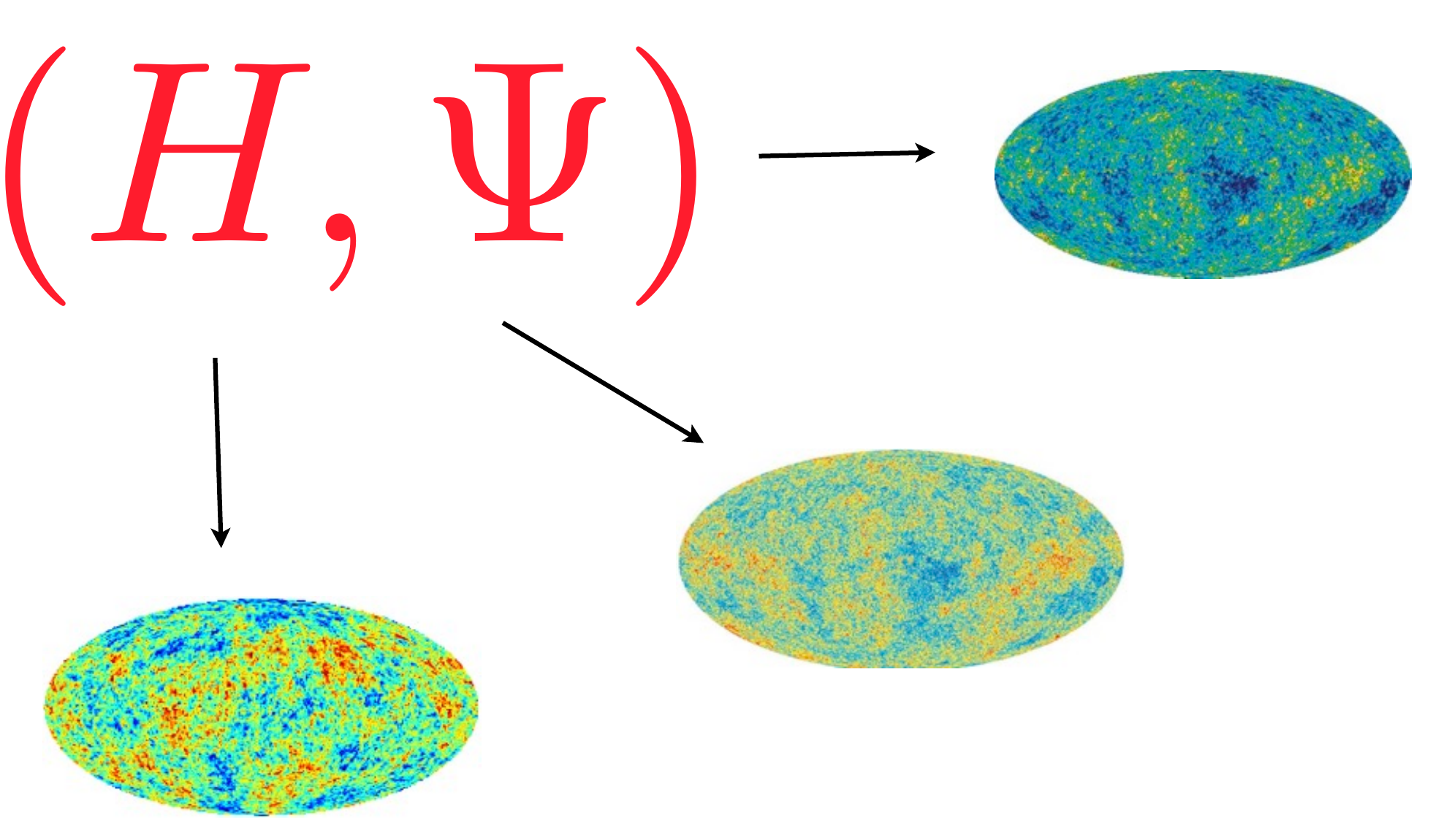}
\caption{There are two theoretical inputs to the process of prediction in quantum cosmology. First, there is a theory of dynamics like string theory denoted here by $H$. Second is the quantum state of the universe denoted here by $\Psi$. 
From these third person probabilities for the individual members decoherent sets of alternative histories of the universe can be derived. From these and data describing our observational situation, first person probabilities predicting the results our observations can be calculated. These include predictions for what CMB maps we will observe.}
\label{cmbbranches}
\end{figure}

{\it Theoretical Inputs: }  Cosmology requires a generalization of usual DH to include spacetime geometry as a quantum mechanical variable \cite{Har95c}. The basic theoretical inputs are a theory of  dynamics including spacetime  $H$-- say some version of string theory --- and  a theory of the quantum state of the universe $\Psi$  --- say Hawking's no-boundary wave function of the universe \cite{HH83}.

{\it Histories:} Our observations of the large scale universe are mostly of properties of its classical history.  The expansion, the amount of inflation, the formation of the fluctuations we see in the cosmic wave background  radiation (CMB), and in the large scale distribution of galaxies today, are all properties of that classical history. Quantum cosmology aims to predict probabilities for these properties by deriving probabilities for alternative classical histories from $(H,\Psi)$ as emphasized in Figure \ref{cmbbranches}.

But classical behavior is not a given in DH. It s a matter of  the quantum probabilities of decoherent sets of appropriately coarse grained histories of geometry and field \cite{Har10}. A system behaves classically when the probabilities are high for histories that exhibit correlations in time by classical deterministic laws such as the Einstein equation. Classical behavior is not built into the quantum mechanics of closed systems as it was in Copenhagen theory but rather an emergent feature of the probabilities supplied by $(H,\Psi)$. 

{\it Observers and Observations:}  Observers and their apparatus are physical  systems within the  universe with only a probability to have evolved in any region of spacetime and, in a very large universe, a probability to be replicated in many regions.  In TSMU the single observer (us) and detector are  physical systems within the universe assumed to exist with unit probability.

{\it 3rd and 1st Person Probabilities:}  In quantum cosmology the  theory $(H,\Psi)$ predicts third person probabilities for the history of the universe that occurs.  From these  probabilities for what we will observe can be predicted.  In very large universes the histories most probable to occur may not be the histories that are most probable to be observed  \cite{HH15b}. The branch dependent  adaptive coarse grainings discussed briefly in Section \ref{firstandthird} are essential for cosmology.  Our observations of the universe are limited --- highly coarse grained. The universe is vast.  We can most efficiently calculate the prediction of theory for the outcomes of our observations by  using sets of histories that  follow what is observed  and coarse grain over  features of the universe that do not affect these observations (e.g. \cite{HHH10b,HH15a}). 

{\it Living in a Superposition:} Just like \eqref{sumbranches} of TSMU the quantum state of the universe is a superposition of the branch state vectors for any decoherent set of alternative classical histories.  Therefore, just like the observer in TSMU, you and I are living in a superposition.  We are all Schr\"odinger cats in Hawking's wave function of the universe.


\renewcommand{\theequation}{\Alph{section}.\arabic{equation}}

\vskip .3in 

\noindent{\bf Acknowledgments:} The author has had the benefit of  discussions with a great many scientists on the work that underlies this paper. The cited papers have those acknowledgments. However, his collaborators on the those papers over a long period of time should be thanked. They are Murray Gell-Mann, Stephen Hawking, Thomas Hertog,  and Mark Srednicki.  Thanks are due to Mark Srednicki for critical readings of the essay and to Simon Saunders for a careful reading of the text. The work  was supported in part by the US NSF grant PHY15-04541.

\end{document}